\documentclass{IEEEtran}
\usepackage{cite}
\usepackage{float}
\usepackage{booktabs}
\usepackage{amsmath,amssymb,amsfonts,multirow}
\usepackage{graphicx}
\usepackage{textcomp,nicefrac}
\usepackage{dcolumn}
\usepackage{bm}
\usepackage{subcaption}
\usepackage{hyperref}
\usepackage{nomencl}
\makenomenclature

\usepackage{newtxtext,newtxmath}
  
\def\BibTeX{{\rm B\kern-.05em{\sc i\kern-.025em b}\kern-.08em
T\kern-.1667em\lower.7ex\hbox{E}\kern-.125emX}}
\begin{document}
\title{The Automatic Calibration Method of the Compton Edge Based on Normalized Cross-correlation and Simulated Annealing Algorithm}

\author{Dehua Kong, Yanbiao Zhang, Zixi Lin, Yehao Qiu, Xiulian Chen\IEEEauthorrefmark{1} and Zhonghai Wang
\thanks{
Manuscript received XX XX, 202X; revised XX XX, 202X; accepted XX XX, 202X. Date of publication XX XX, 202X; date of current version XX XX, 202X. 
Corresponding author: Zhonghai Wang (e-mail: zhonghaiwang@scu.edu.cn). 

All authors are with the Department of Physics, Sichuan University, Chengdu 610065, China. 
E-mails: dehuak@foxmail.com (D. Kong); 2979324403@qq.com (Y. Zhang); 13405930952@163.com (Z. Lin); 13879008264@163.com (Y. Qiu); cxl@scu.edu.cn (X. Chen); zhonghaiwang@scu.edu.cn (Z. Wang).
}
}

\maketitle

\begin{abstract}
Accurate energy channel calibration in scintillation detectors is essential for reliable radiation detection across nuclear physics, medical imaging, and environmental monitoring. Organic scintillators like BC408 and EJ309 lack full-energy peaks, making their Compton edge a critical calibration alternative 
where traditional peak methods fail. Existing Compton edge identification techniques—Gaussian fitting for the 50\%~70\% amplitude point, first derivative minimum detection, and Monte Carlo simulation—suffer significant degradation from low count rates, spectral overlap, and subjective interval selection. 
For the first time, we propose an automated calibration procedure based on Normalized Cross-Correlation (NCC), Simulated Annealing (SA), and a convolutional response 
model to address these issues. This method automates the selection of the Compton edge interval through NCC-based matching, utilizes SA for global parameter optimization, and then employs a convolutional model for precise matching. Experiments involving the irradiation of organic scintillators (BC408, EJ309) 
and inorganic scintillators (NaI:Tl, LaBr$_{3}$:Ce) with $^{137}$\text{Cs}, $^{22}$\text{Na}, $^{54}$\text{Mn}, and $^{60}$\text{Co} radiation sources demonstrate that this method achieves accuracy commensurate with full-energy peak calibration method (cosine similarity>99.999\%) and exhibits superior stability compared to the two traditional methods.
 In the extreme cases of spectral overlap and low count rate, the average errors of this method compared with the two traditional methods have been reduced by 80\%, 85\%, 44\% and 66\% respectively.
This work advances detector calibration and offers a scalable, automated solution for high-energy experiments and portable devices.\end{abstract}


\section{Introduction}
\label{sec:introduction}
\IEEEPARstart{S}{cintillation} Detectors play a critical role in radiation detection, forming the backbone of applications across nuclear physics, medical diagnostics, industrial monitoring, and radiation protection~\cite{Wang_2023}. In the field of radiation detection, the performance of scintillator detectors, 
as the core detection element, is directly related to the reliability of nuclear physics research, medical diagnosis, industrial monitoring, and radiation protection applications. The accuracy of energy calibration is the key factor to ensure the performance of the detector. For inorganic scintillator detectors,
the traditional calibration method of full-energy peak has been developed, but this method faces major challenges in plastic scintillator detectors. Because the plastic scintillator material has the characteristics of low atomic number, its interaction mechanism is dominated by Compton scattering rather 
than photoelectric effect, which makes it difficult to show the full energy peak in the energy spectrum, leading to the failure of the traditional calibration method.

To address the energy calibration challenges in plastic scintillators, researchers have developed calibration methods based on the Compton edge. However, in practical applications, due to the Gaussian broadening effect of the detector energy response, the Compton edge exhibits a gradual descent rather than the
theoretically predicted sharp drop. Current mainstream approaches encompass three strategies: The first method employs Gaussian function fitting to the spectral descent slope, defining the channel position at 50\% of the maximum amplitude as the Compton edge~\cite{SongYun2023}. Yet this approach is highly sensitive to spectral
noise, and the assumed ideal Gaussian distribution significantly deviates from the observed asymmetric broadening, leading to systematic errors in edge localization. The second method calculates the first derivative of the smoothed measured energy spectrum, identifying the channel corresponding to the derivative 
minimum as the Compton edge~\cite{Zhang2021}. While effective in suppressing noise-induced fluctuations, it remains vulnerable to noise interference at low count rates, particularly for low-Z scintillators with overlapping spectral features, where manual interval selection substantially impacts accuracy. The third method utilizes
Monte Carlo simulations to generate theoretical energy spectra across varying energy resolutions (0.5\%--20\%), and determines the Compton edge position by selecting the optimal theoretical spectrum through $\chi^2$ minimization from an extensive parameter space (40 resolutions $\times$ 120 cutoff
channels $\times$ 60 gains). Although this method significantly mitigates noise and resolution broadening effects via precise spectral matching, it suffers from high computational complexity, requires source-specific simulations, and exhibits rapid performance degradation when actual energy resolution exceeds 
the pre-simulated range, thus limiting its generalizability~\cite{SEO2011S108}.

In this work, we introduce an automatic calibration method grounded in normalized cross-correlation (NCC), simulated annealing (SA), and a convolutional model, hereinafter referred to as the ``NCC-SA method''. The convolutional model incorporates the Gaussian broadening effect, enabling precise matching between 
theoretical and experimental spectra. By utilizing NCC, we achieve accurate and objective Compton edge identification without the need for manual interval selection. Additionally, SA facilitates global optimization of model parameters, ensuring robustness against local minima that traditional gradient-based 
algorithms may fail to escape. The NCC-SA method is effective for a wide range of scintillation detectors, including BC408, EJ309, NaI:Tl, and LaBr$_3$:Ce, as well as various radiation sources. Experimental results for inorganic scintillators (NaI:Tl and LaBr$_3$:Ce) demonstrate its comparable accuracy to the 
full-energy peak calibration method. For organic scintillators (BC408 and EJ309), the experimental results showcase its superior stability and robustness compared to two traditional Compton edge calibration methods.
\section{Method of Energy Calibration}
The Compton edge is a characteristic feature formed when gamma rays undergo Compton scattering with electrons in the detector material. When the scattering angle is 180 degrees (i.e., the maximum scattering angle), the recoil electron acquires the maximum energy $E_{max}$, and the corresponding pulse amplitude of the energy spectrum at this point is the Compton edge. Based on the conservation of energy and momentum, the energy calculation formula can be derived as:

\begin{equation}
    E_{\max }=\frac{E_\gamma}{1+2 E_\gamma /\left(m_e c^2\right)}
    \end{equation}

    The differential scattering cross-section is described by the Klein-Nishina formula:
    \begin{equation}
        \begin{aligned}
        & \frac{\mathrm{d} \sigma_{c,e}}{\mathrm{~d} \Omega}=\mathrm{r}_0^2\left(\frac{1}{1+\alpha(1-\cos \theta)}\right)^2\left(\frac{1+\cos ^2 \theta}{2}\right) \\
        & \times\left(1+\frac{\alpha^2(1-\cos \theta)^2}{\left(1+\cos ^2 \theta\right)[1+\alpha(1-\cos \theta)]}\right)
        \end{aligned}
        \end{equation}
        
        Where $\frac{d \sigma_{c,e}}{d \Omega}$ is the Compton scattering differential cross-section, representing the probability of a photon being scattered into a solid angle of unit size when incident on a medium containing a single electron per unit area. $\alpha=\frac{E_\gamma}{m_e c^2}$, is the dimensionless ratio of the incident photon energy to the electron rest mass energy, and $\mathrm{r}_0$ is the classical electron radius, approximately equal to $2.818 \times 10^{-15} \mathrm{~m}$.

        By substituting the relevant equations, the recoil electron's energy differential cross-section reveals an asymmetric distribution, characterized by a sharp drop at the Compton edge, as depicted in the Fig.\ref{fig:differential}:

        \begin{figure}[t]
            \centerline{\includegraphics[width=3.5in]{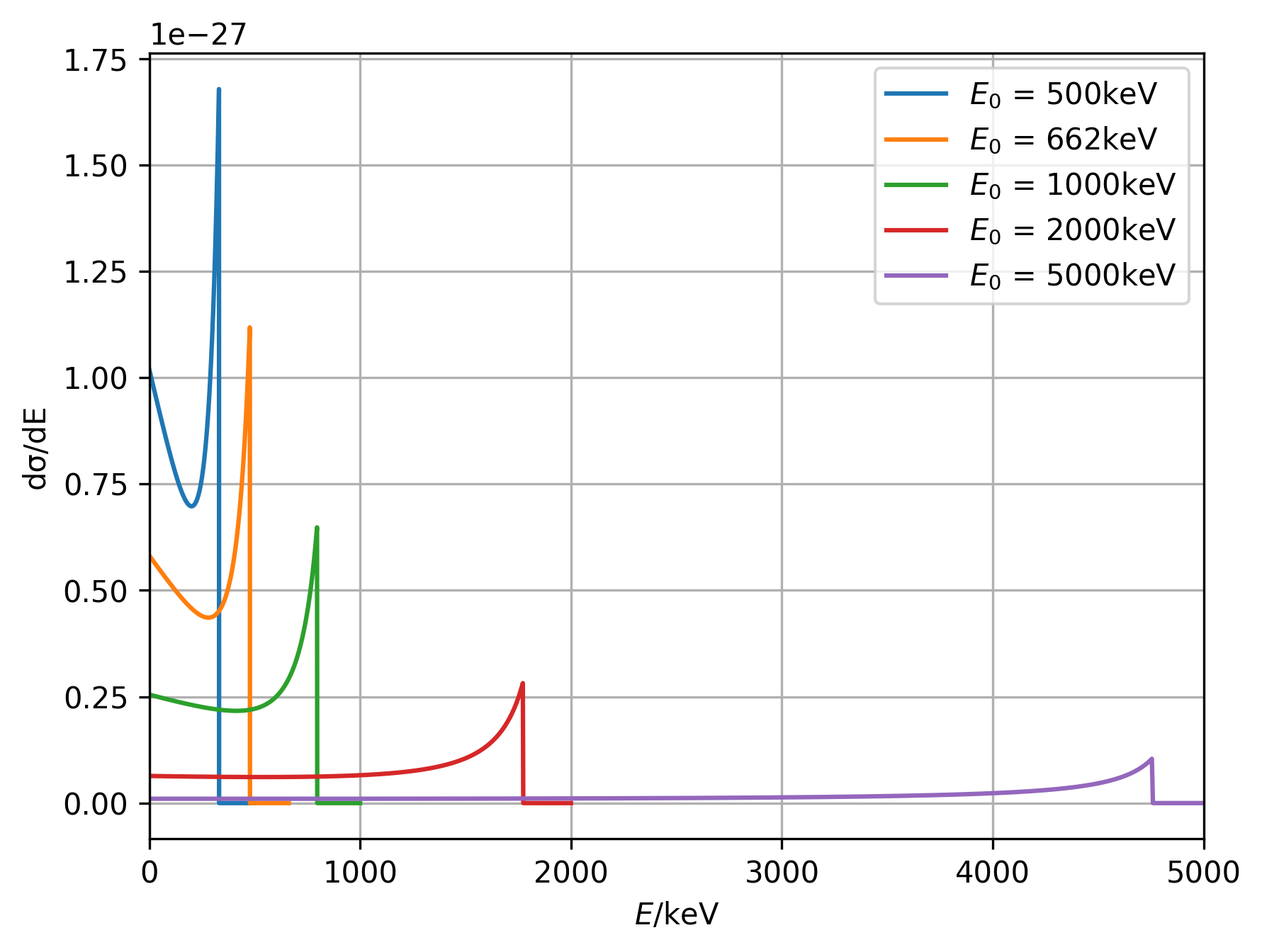}}
            \caption{Differential Cross Section for Recoil Electron Energy. Here $E_0$ is the incident photon energy.}
            \label{fig:differential}
        \end{figure}

\subsection{Convolution Model and Simulated Annealing}        
Experimental observations show that the Compton edge measured by the detector differs somewhat from theoretical predictions, and that the edge is not characterized by an ideal vertical form. This discrepancy mainly stems from the energy spectrum broadening effect caused by several factors, among which the intrinsic properties of the scintillator detector introduce significant Gaussian broadening. The specific influencing factors are shown in Table~\ref{tab1}\cite{Lecoq2017}:

\begin{table}[ht]
    \caption{Broadening Source of the Compton Edge}
    \label{tab1}
    \centering
    \begin{tabular}{p{120pt}p{120pt}}
        \toprule
        \textbf{Broadening Source} & \textbf{Formula} \\
        \midrule
        Fluctuation of light yield & $\sigma_{\mathrm{LY}} = \sqrt{N_{\mathrm{ph}}} \cdot \mathrm{LY}_0$ \\
        SiPM gain fluctuation & $\sigma_{\mathrm{gain}} = \mathrm{ENF} \cdot \sqrt{N_{\mathrm{pe}}}$ \\
        Electronic thermal noise & $\sigma_{\mathrm{elec}} = \sqrt{(4kT + 2qIR) R \Delta f}$ \\
        \bottomrule
    \end{tabular}
\end{table}

The resultant broadening is the sum of squares of the independent terms\cite{Knoll2010Radiation}:
\begin{equation}
    \sigma_{\mathrm{total}}=\sqrt{\sigma_{\mathrm{LY}}^2+\sigma_{\mathrm{gain}}^2+\sigma_{\mathrm{elec}}^2}
\end{equation}

In order to eliminate this effect and obtain a true spectrum (ideal spectrum), the measured spectrum is usually back-convoluted according to the resolution function of the instrument\cite{Paatero1974-PAADIC}. The effect of the function of common physical devices such as amplifiers, filters, and optical instruments on the resolution can be summarized as the following systematic process:
\begin{equation}
    \mathrm{input~x}(t)\to {device~h}(t)\to {output~y}(t)
\end{equation}

The output $y(t)$ depends not only on the input $x(t)$ but also on the detector. In ground calibration experiments, the scintillator detector can be regarded as a linear time-invariant system due to the approximate invariance of the experimental environment, and the output $h(t)$ follows the linearity and time invariance:

\begin{equation}
\begin{cases}
\alpha x_{1}(t)+\alpha_2x_2(t)\to\alpha_1 y_1(t)+\alpha_2 y_2(t),&\text{If} x_{1,2}(t) \to y_{1,2}(t) \\
x(t-\tau) \to y(t-\tau),&\text{If} x(t) \to y(t)
\end{cases}
\end{equation}
    
When the input of the system is $\delta\left(t\right)$, the output of the system is $h\left(t\right)$, and due to the nature of $\delta\left(t\right)$, any function can be represented by the convolution of the $\delta\left(t\right)$ function:
\begin{equation}f(t)=\int_{-\infty}^{+\infty}f(\tau)\delta(t-\tau)d\tau\end{equation}

And because this system is a linear time-invariant system, the output of the system can be considered as a linear superposition of the output of countless $\delta\left (t\right)$ input signals, linear superposition coefficients for $f\left (\tau\right)$. Therefore, the output signal obtained from all input signals passing through the system can be expressed as the convolution of the input signal $x(t)$ with the response of the system $\delta\left(t\right)$:

\begin{equation}y(t)=\int_{-\infty}^{+\infty}x(\tau)h(t-\tau)d\tau\end{equation}

For the scintillator detector, considering the Gaussian broadening, we consider the response function for each single-photon incident signal $\delta\left(t\right)$ to be a Gaussian function \(g(E)=\frac{A}{\sigma\sqrt{2\pi}} \mathrm{e}^{-\frac{(E-\mu)^2}{2\sigma^2}}\), and so we can Considering that the response function of the detector is the Gaussian function, the energy spectrum function obtained by the detector is the convolution of the energy space differential scattering cross section of the recoil electron with the Gaussian nucleus:

\begin{equation}S_{\mathrm{model}}(E)=\int_{-\infty}^{+\infty}\frac{d\sigma}{dE_{e}}\left(E_{eff}=E_{e}\cdot\frac{B}{E_{max}}\right)\cdot G(E_{e})dE_{e}\end{equation}

Where the Gaussian kernel:  
\begin{equation}G(E_e)=\frac{1}{\sqrt{2\pi}\sigma}\exp\left(-\frac{E_e^2}{2\sigma^2}\right)\end{equation}

The differential scattering cross section is:
\begin{equation}
    \begin{aligned}
    \frac{d\sigma}{dE_{e}} &= \frac{A \pi r_{0}^{2}}{\alpha^{2} m_{0} c^{2}} \cdot \left[ 2 + \left( \frac{E_{e \mathrm{ff}}}{h\nu - E_{e \mathrm{ff}}} \right)^{2} \right. \\
    & \quad \left. \times \left( \frac{1}{\alpha^{2}} - \frac{2(h\nu - E_{e \mathrm{ff}})}{\alpha E_{e \mathrm{ff}}} + \frac{h\nu - E_{e \mathrm{ff}}}{h\nu} \right) \right]
    \end{aligned}
\end{equation}

where $A$, $B$ and $\sigma$ are all parameters to be tuned. $A$ is the amplitude scaling factor of the cross-section (regulating the cross-section strength);$B$ is the Compton edge energy channel parameter; and $\sigma$ is the detector resolution parameter (standard deviation of Gaussian spread).

To address overlapping Compton edge spectra from multiple gamma-ray energies (e.g., $^{60}\text{Co}$ and $^{22}\text{Na}$), we implement a parameter optimization scheme with the following parameters:

\begin{equation}
\Theta = \left\{ A_1, A_2, B_1, \sigma_1, \sigma_2, D \right\}
\label{eq:params}
\end{equation}

\nomenclature{$A_1, A_2$}{Step height parameters for 1st and 2nd Compton edge}
\nomenclature{$B_1$}{Baseline shift parameter for 1st edge}
\nomenclature{$C_1, C_2$}{Slope parameters for 1st and 2nd edge}
\nomenclature{$D$}{Energy scaling coefficient}

Parameters $A_1$, $A_2$, $B_1$, $\sigma_1$, and $\sigma_2$ retain their physical interpretations as defined above,with subscripts denoting the respective Compton edge (1: lower energy, 2: higher energy). The critical innovation for handling energy separation is expressed as:

\begin{equation}
\text{CE}_{\text{ch},2} = B_1 + D \cdot \Delta E
\label{eq:edge2}
\end{equation}

\noindent where:
\begin{itemize}
  \item $\text{CE}_{\text{ch},2}$ is the Compton edge channel position for the higher-energy gamma ray
  \item $\Delta E = E_2 - E_1$ is the energy difference between gamma lines
  \item $D$ represents the detector's energy-to-channel conversion coefficient
\end{itemize}

The composite spectrum $S_{\text{comp}}(ch)$ for overlapping edges is then obtained by superposition of the individually optimized edge functions:

\begin{equation}
S_{\text{comp}}(ch) = f_1(ch; A_1, B_1, C_1) + f_2(ch; A_2, \text{CE}_{\text{ch},2}, C_2)
\label{eq:superposition}
\end{equation}

\noindent where $f_1(\cdot)$ and $f_2(\cdot)$ are the parameterized Compton edge functions for the respective gamma energies.

Based on the above analysis, it can be seen that the key to accurately determine the position of the Compton edge is to eliminate the influence of the detector response function (Gaussian function) on the pulse amplitude spectrum. To address the above problems, this study proposes a Gaussian spreading elimination algorithm based on convolutional operation.

In the parameter optimization process, we define the loss function as follows:
\begin{equation}loss_{min}(E_i)=\sum_{i=1}^N[S_1(E_i)-S_{\mathrm{model}}(E_i;\vec{p})]^2\end{equation}

Where $E_i$ denotes the $\mathrm{i}$th energy point, $S_1\left(E_i\right)$ is the experimental measurement, $S_{\mathrm{model}}\left(E_i;\vec{p}\right)$ is the model prediction, and $\vec{p}$ is the parameter vector to be optimized. The combination of parameters that minimizes the loss function through iterative optimization is the optimal solution.

In the multi-parameter optimization model, the simulated annealing (SA) algorithm is employed to address the issue of local optima. The SA process initiates with parameter vector $\mathbf{x}_0$ at $T_{\text{init}} = 2000$ and cooling coefficient $\alpha = 0.95$. During each temperature stage $T_k$, the algorithm executes 100 Markov chain iterations by generating new states $\mathbf{x}_{\text{new}} = \mathbf{x}_{\text{current}} + \Delta\mathbf{x}$ where $\Delta x_i \sim \mathcal{U}(-0.1\delta_i, 0.1\delta_i)$ with $\delta_i$ defining the parameter search range. State transitions follow the Metropolis criterion: new configurations are accepted with probability $P = \exp(-\Delta E/T_k)$ where $\Delta E = E(\mathbf{x}_{\text{new}}) - E(\mathbf{x}_{\text{current}})$ and $E(\mathbf{x})$ represents the NCC-based matching error objective function. After completing Markov chains at current temperature, the system cools geometrically via $T_{k+1} = \alpha \cdot T_k$ ($\alpha \in [0.85, 0.99]$). This thermodynamic cycle iterates until meeting dual termination criteria: $T_k < 10^{-6}$ or energy stabilization ($|\Delta E_{\text{mean}}| < 0.001$ over 10 consecutive cycles), ensuring progressive convergence toward the global optimum while escaping local minima.

\subsection{Normalized Cross-correlation}
In complex gamma energy spectral analysis, the determination of Compton edge energy channel ranges traditionally relies mainly on manual calibration, which is not only inefficient but also prone to introduce subjective errors. In order to solve this problem, this study adopts an automatic region selection method based on the Pearson correlation coefficient.The formula of the Pearson correlation coefficient is as follows:

\begin{equation}\rho(\vec{p})=\frac{\sum_{i=1}^N[S_1-\overline{S_1}][S_2-\overline{S_2}]}{\sqrt{\sum_{i=1}^N[S_1-\overline{S_1}]^2\sum_{i=1}^N[S_2-\overline{S_2}]^2}}\end{equation}

Where $S_1$ denotes the measured data and $S_2$ denotes the characterized data. the Pearson correlation coefficient ($\rho$) is an important indicator to measure the degree of linear correlation between the two sets of data, and its value ranges from [-1, 1]. When $\rho=1$, it indicates that the two sets of data show a completely positive correlation and can be completely overlapped by point-to-point translation, i.e., the waveforms are completely similar; on the contrary, when $\rho=-1$, it indicates that the two sets of data are not similar at all.

The Compton edge localization is performed by comparing experimental pulse amplitude spectra with theoretical predictions. Assuming the measured spectrum represents a convolution of the theoretical response with a Gaussian function, we first generate 100 reference points ($S_2$) by uniformly sampling the predicted lineshape from the Compton plateau to the edge minimum. For experimental data, each non-zero point serves as a potential edge start position, with subsequent points forming test intervals that are interpolated to 100 points ($S_1$). The interval showing maximum Pearson correlation ($\rho$) between $S_1$ and $S_2$ is identified as the optimal Compton edge region.

\begin{figure}[t]
    \centerline{\includegraphics[width=3.5in]{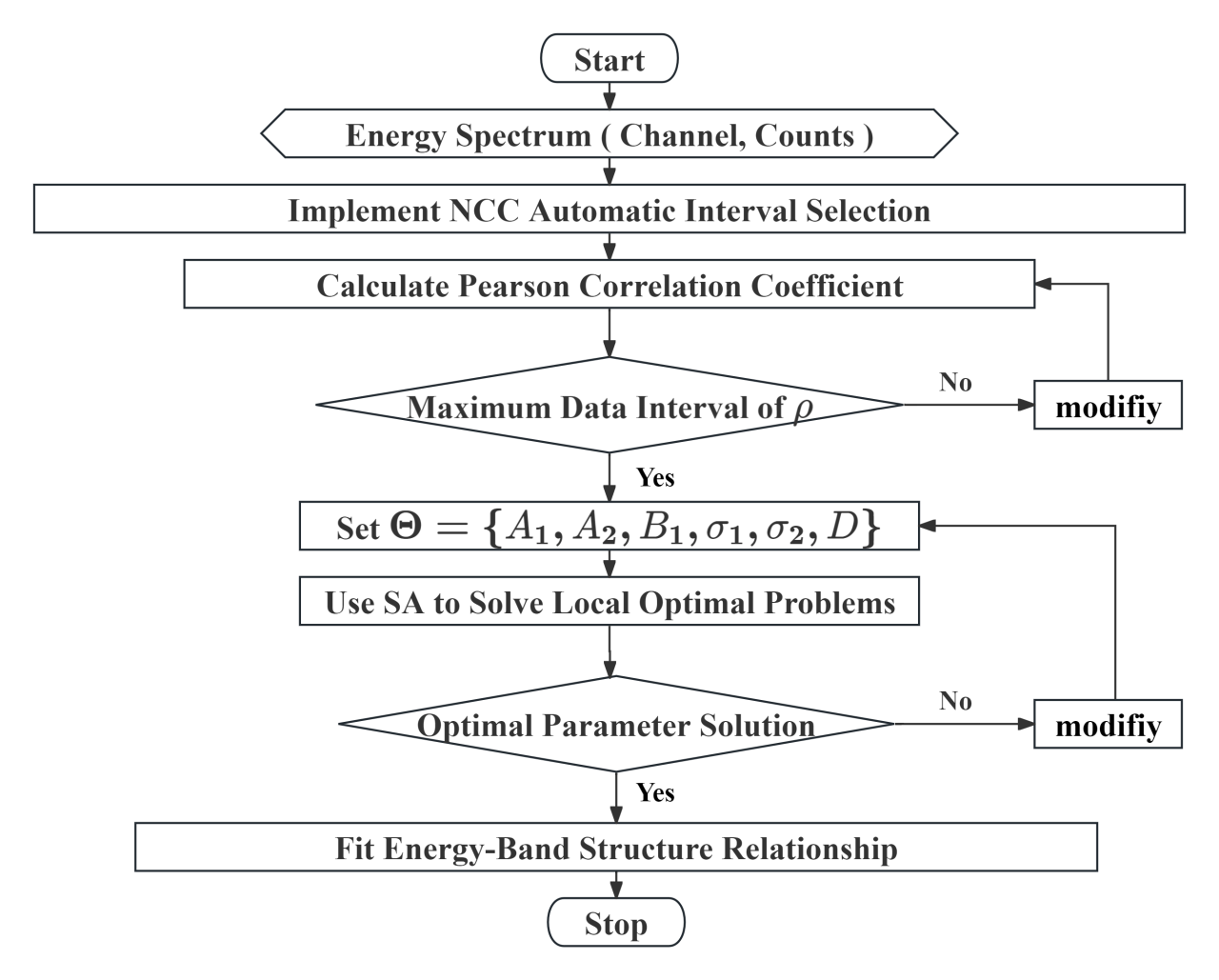}}
    \caption{the specific flow of the NCC-SA method}
    \label{fig:The specific flow of the NCC-SA method}
\end{figure}

The specific flow of the NCC-SA method is shown in the Fig.\ref{fig:The specific flow of the NCC-SA method}. The method first uses the NCC method to automatically select the Compton edge interval, and then uses the SA algorithm to optimize the parameters of the convolution model. Finally, the energy channel relationship is obtained by fitting the Compton edge position.

\subsection{Experimental Validation Methodology}
To validate the NCC-SA calibration accuracy, inorganic scintillators (LaBr$_3$:Ce, NaI:Tl) were irradiated with $^{137}$Cs, $^{60}$Co, $^{54}$Mn, and $^{22}$Na sources. The resulting calibration was benchmarked against standard Gaussian peak fitting performed according to GB/T 7167-2008 (equivalent to IEC 60973:2008 \textit{Test methods for germanium gamma-ray detectors}). For stability assessment, organic scintillators (BC408, EJ309) lacking distinct full-energy peaks underwent identical irradiation protocols. Ten independent pulse-height spectra measurements per scintillator-source combination yielded 80 datasets, with $^{60}$Co spectra exhibiting Compton-edge overlap and $^{22}$Na spectra acquired at $\leq$10,000 counts to simulate low-efficiency environments. Calibration curves derived from NCC-SA and two conventional methods—Gaussian full-width-at-half-maximum (FWHM) fitting and differential minimum detection—were generated by averaging each scintillator-type data. Experimental energies were reconstructed from the energy-channel relationships and compared against theoretical values to calculate relative errors with error bars representing standard deviations across measurements.

\section{Results}

\begin{figure*}[htbp]
    \centering
    \begin{subfigure}{0.48\textwidth}
        \centering
        \includegraphics[width=\linewidth]{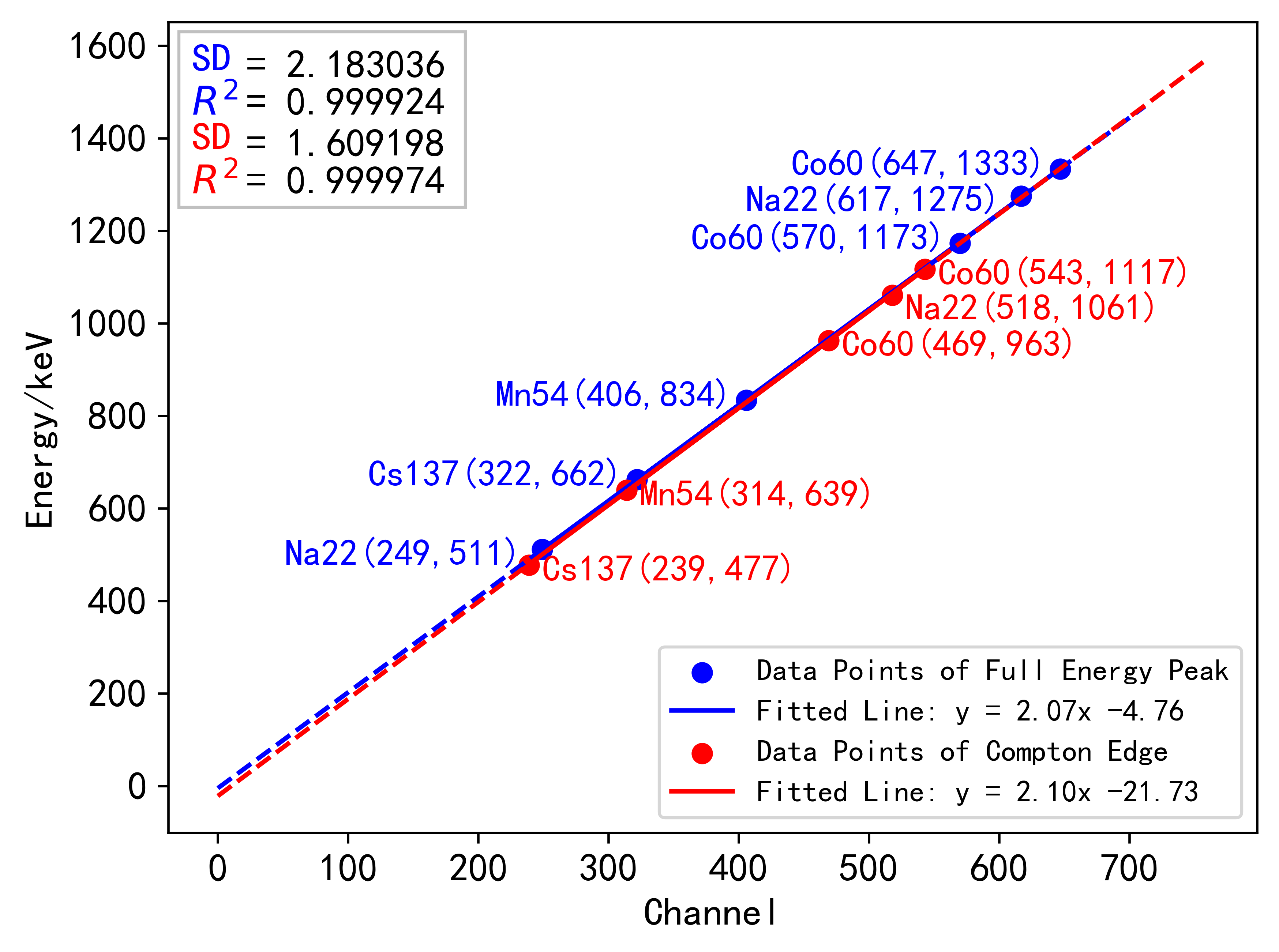}
        \caption{Calibration results for LaBr$_3$:Ce detector.}
        \label{fig:5}
    \end{subfigure}
    \hfill
    \begin{subfigure}{0.48\textwidth}
        \centering
        \includegraphics[width=\linewidth]{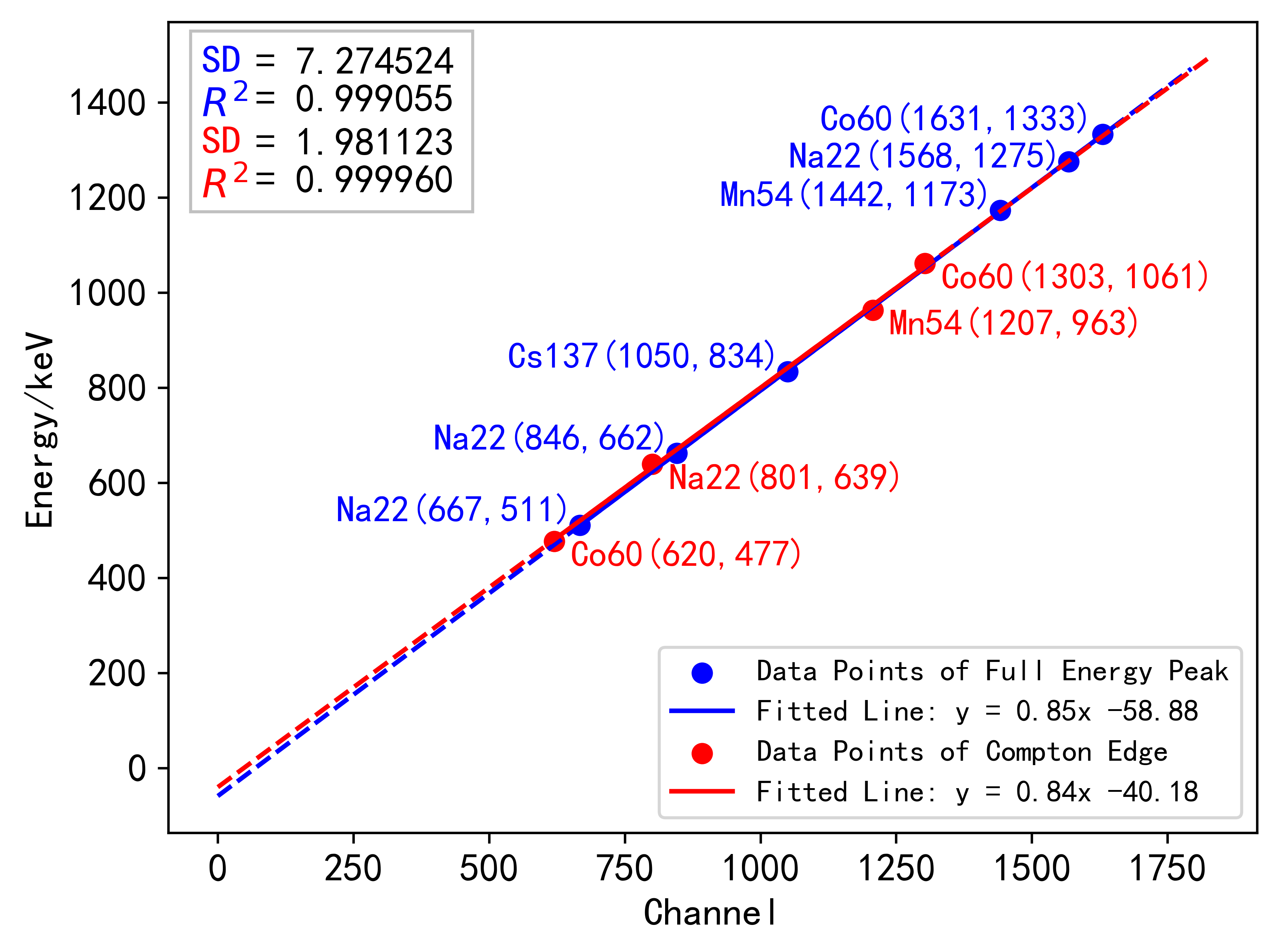}
        \caption{Calibration results for NaI:Tl detector.}
        \label{fig:6}
    \end{subfigure}
    \caption{Energy calibration results comparing NCC-SA (red) and full-energy peak (blue) methods. (a) LaBr$_3$ detector spectrum with corresponding calibrations. (b) NaI detector spectrum with corresponding calibrations. Error metrics include standard deviation (SD) and coefficient of determination ($R^2$).}
    \label{fig:comparison_peak}
\end{figure*}

To evaluate the accuracy of the NCC-SA method, calibration was performed using cerium bromide (LaBr$_{3}$:Ce) and sodium iodide (NaI:Tl) detectors, both of which can produce distinct full-energy peaks. The NCC-SA results were compared with the calibration results obtained by fitting the Gaussian peak according to GB/T 7167---2008 \textit{Test methods for germanium gamma-ray detectors}. As shown in Fig.~\ref{fig:comparison_peak}, the energy-channel relationship obtained by the two methods is very consistent. Through calculation, it can be seen that the cosine similarity of the energy-channel relationship of the LaBr$_{3}$:Ce detector reaches 99.9993\%, while the cosine similarity of the NaI:Tl detector is as high as 99.9999\%, indicating that the accuracy of the NCC-SA method is comparable to that of the full energy peak fitting method. Crucially, the standard deviation associated with the NCC-SA calibration consistently remains below that of the full energy peak fitting method, indicating higher precision.

\begin{figure*}[htbp]
    \centering
    \begin{subfigure}{0.48\textwidth}
        \centering
        \includegraphics[width=\linewidth]{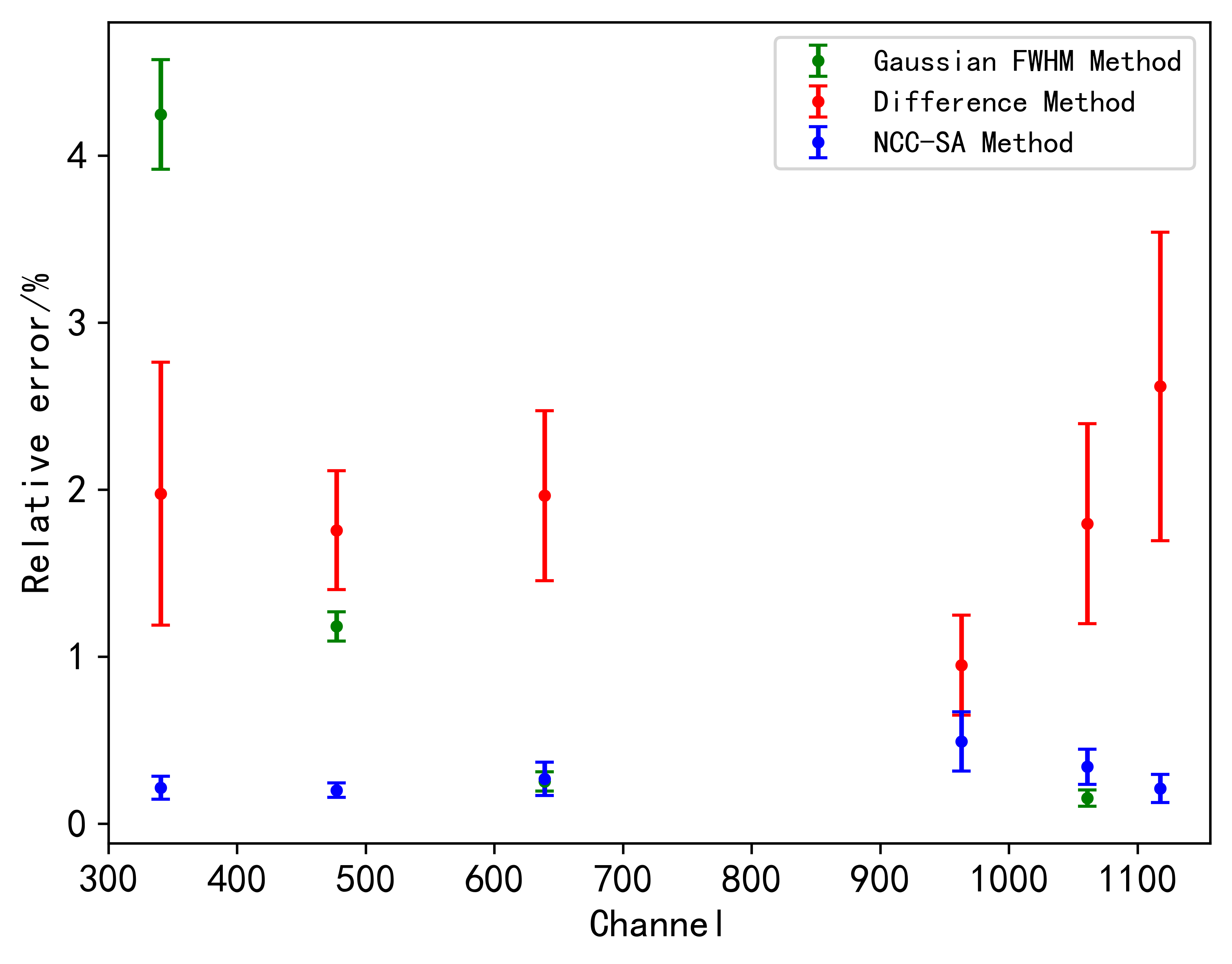}
        \caption{Comparison of calibration errors for BC408 scintillator.}
        \label{fig:error_bc408}
    \end{subfigure}
    \hfill
    \begin{subfigure}{0.48\textwidth}
        \centering
        \includegraphics[width=\linewidth]{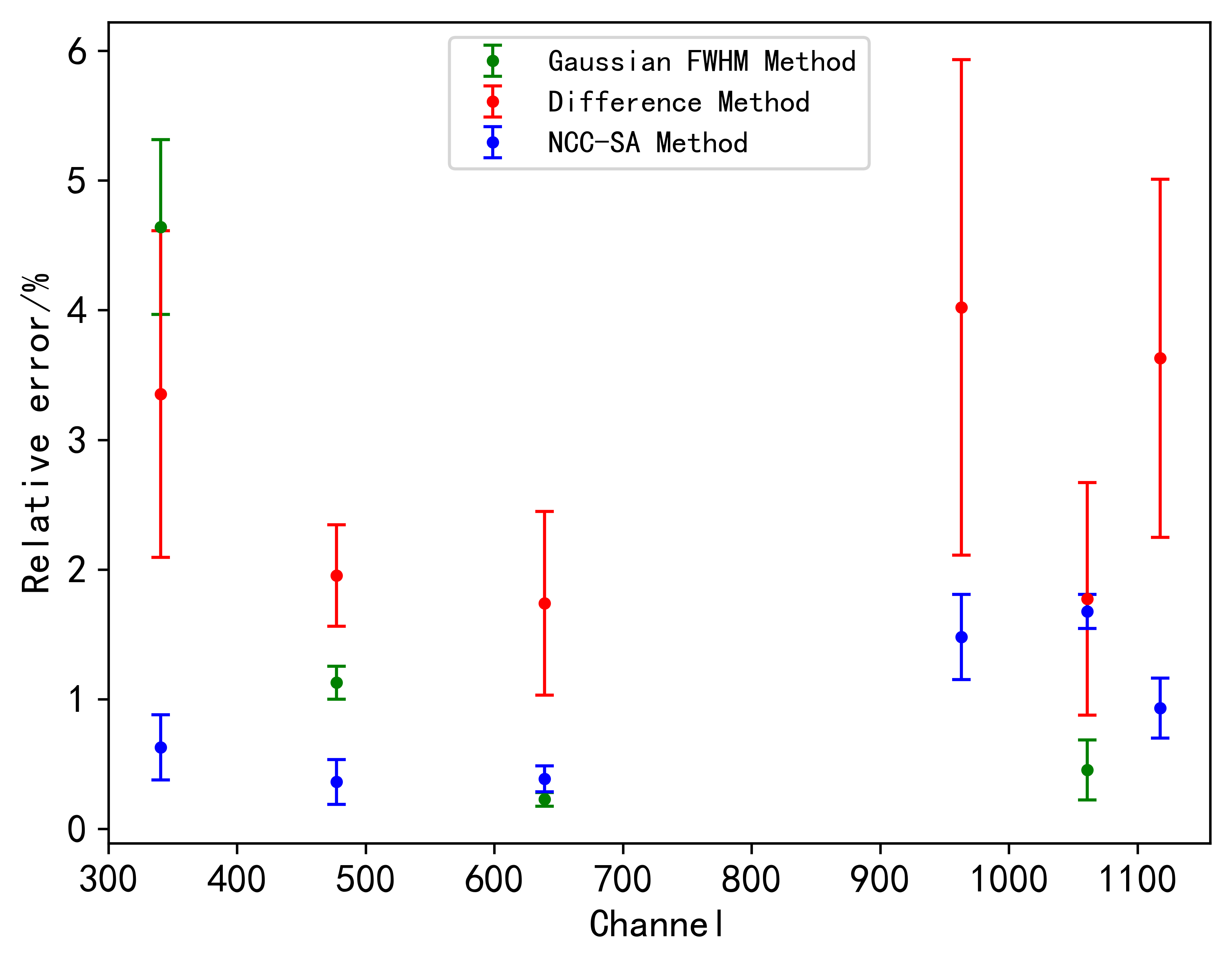}
        \caption{Comparison of calibration errors for EJ309 scintillator.}
        \label{fig:error_ej309}
    \end{subfigure}
    \caption{Relative errors and standard deviations of Compton edge calibration for (a) BC408 and (b) EJ309 scintillators using NCC-SA Method, Gaussian FWHM Method, and Difference Method.}
    \label{fig:error_comparison}
\end{figure*}

Furthermore, we performed energy-channel calibrations on BC408 and EJ309 organic scintillators (devoid of full-energy peaks) using the NCC-SA method and two conventional methods (Gaussian FWHM Method and Difference Method) to validate the stability and robustness of NCC-SA. Ten independent measurements were conducted for each detector exposed to $^{137}$Cs, $^{60}$Co, $^{54}$Mn, and $^{22}$Na sources, yielding 80 pulse amplitude spectra. The $^{60}$Co spectra exhibited overlapping Compton edges, while $^{54}$Mn spectra were deliberately acquired at count rates below 10,000 to simulate low-flux conditions. The Compton edge energies derived from the energy-channel relationships calibrated by the three methods are compared with theoretical values in Fig.~\ref{fig:error_comparison}. Error bars indicate the standard deviation of channel positions (reflecting data dispersion), and data points represent the relative error between the mean channel position and theoretical value. Results demonstrate that both conventional methods exhibit significantly larger error bars and data point deviations than NCC-SA. Notably, the Gaussian FWHM method failed to resolve overlapping edges in $^{60}$Co and $^{22}$Na, while the Difference Method showed substantial errors for $^{60}$Co overlapping edges, $^{22}$Na-511keV low-count regions, and $^{54}$Mn low-flux spectra. In contrast, NCC-SA maintained superior performance across all extreme conditions: it reduced mean errors by 80\% and 85\% compared to differential and Gaussian methods in BC408, and by 44\% and 66\% in EJ309, demonstrating exceptional stability.

\begin{figure*}[htbp]
    \centering
    \includegraphics[width=\textwidth]{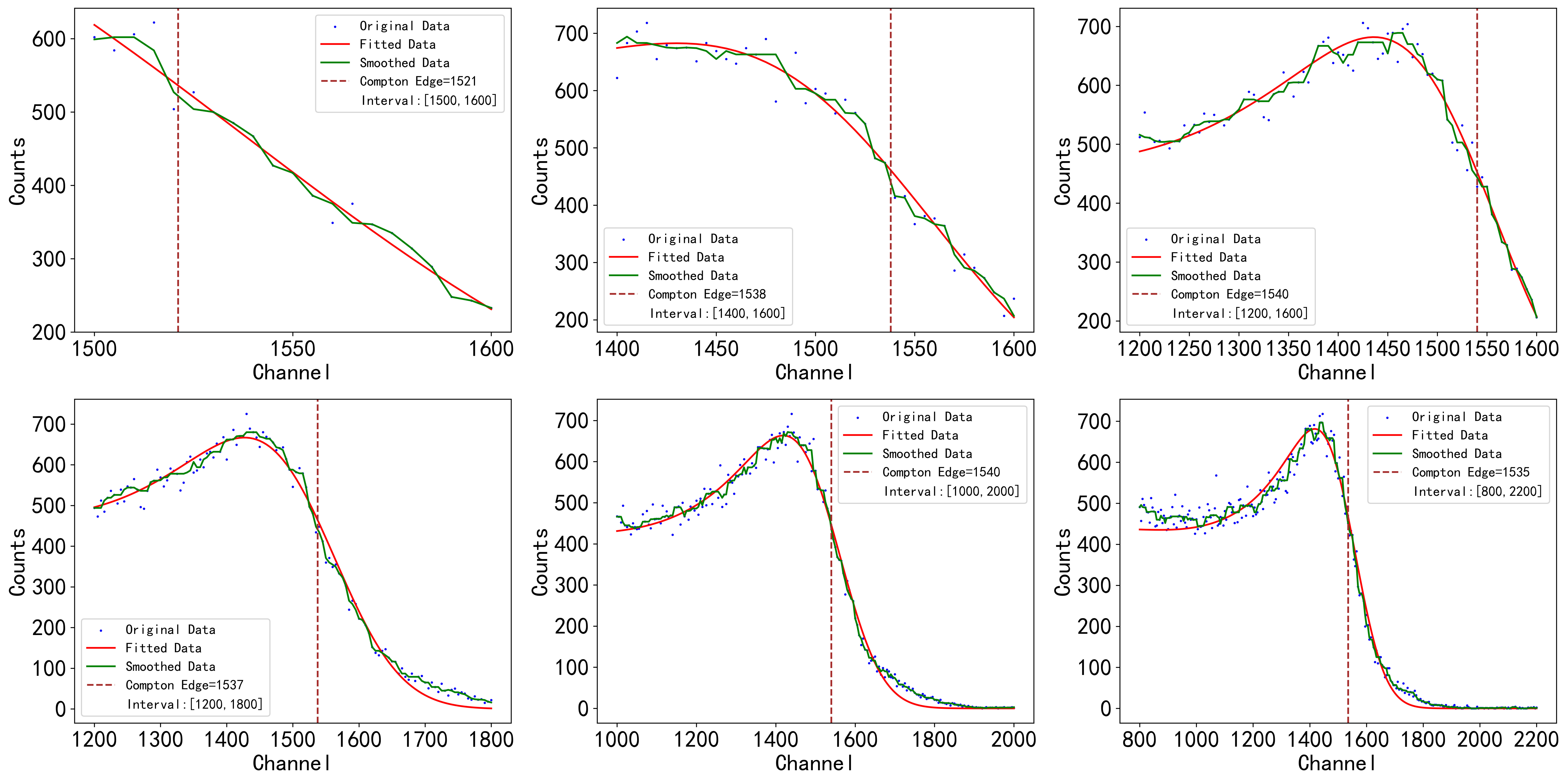}
    \caption{Robustness of the NCC-SA calibration results for BC408 scintillator (\textsuperscript{137}Cs source) under different fitting interval selections. The results remain highly consistent, demonstrating reduced sensitivity to operator-dependent parameter choices.}
    \label{fig:interval_robustness}
\end{figure*}

The NCC-SA method exhibits reduced sensitivity to operator choices, primarily the selection of the fitting interval. This robustness was tested by calibrating EJ309 spectra (\textsuperscript{137}Cs source) using fitting intervals of varying lengths. As illustrated in Fig.~\ref{fig:interval_robustness}, the calibration results remained highly consistent across the different intervals, confirming the method's robustness against this potential source of variability.

\section{Conclusion}

This study proposes a novel automatic calibration method for scintillation detectors integrating normalized cross-correlation (NCC) and simulated annealing (SA) with a convolutional response model. The framework overcomes conventional limitations in Compton edge identification by automating interval selection through NCC matching, optimizing parameters globally via SA, and suppressing spectral broadening via convolutional modeling. Experimental validation on inorganic scintillators (NaI:Tl, LaBr$_3$:Ce) demonstrates accuracy equivalent to full-energy-peak calibration, evidenced by a cosine similarity of 99.999\%. For organic scintillators (BC408, EJ309) under extreme conditions—including low count rates (down to 10k counts), spectral overlaps ($^{60}$Co/$^{22}$Na), and manual interval selection—the method achieves mean error reductions of 80\%/85\% (BC408) and 44\%/66\% (EJ309) compared to conventional approaches. These quantitative results confirm superior stability and robustness, providing a reliable automated solution for radiation detection applications.

\section*{Acknowledgment}
The author gratefully acknowledges Associate Professor Wang Zhonghai for insightful discussions and guidance throughout this research. Appreciation also extends to Zhao Qianru and Mao Zhiyuan for providing experimental equipment and facilities. This work was supported by Sichuan University Undergraduate Students' Innovation and Entrepreneurship Training Program under Grant C2024128729.
\section{References}
\bibliographystyle{unsrt}
\bibliography{ComptonEdge.bib}

\begin{thebibliography}{1}

\bibitem{Wang_2023}
Zhehui Wang, Christophe Dujardin, Matthew~S. Freeman, Amanda~E. Gehring, James~F. Hunter, Paul Lecoq, Wei Liu, Charles~L. Melcher, C.~L. Morris, Martin Nikl, Ghanshyam Pilania, Reeju Pokharel, Daniel~G. Robertson, Daniel~J. Rutstrom, Sky~K. Sjue, Anton~S. Tremsin, S.~A. Watson, Brenden~W. Wiggins, Nicola~M. Winch, and Mariya Zhuravleva.
\newblock Needs, trends, and advances in scintillators for radiographic imaging and tomography.
\newblock {\em IEEE Transactions on Nuclear Science}, 70(7):1244–1280, July 2023.

\bibitem{SongYun2023}
Song Yun, Zuo Jingxin, Liang Yongfei, Han Bing, Bai Lixin, and Yang Chaowen.
\newblock Neutron/$\gamma$ discrimination capability testing of self-developed plastic scintillator detector.
\newblock {\em Nuclear Technology}, 46(3), 2023.

\bibitem{Zhang2021}
Shuangjiao Zhang.
\newblock {\em Energy Spectrum Measurement and Characteristics Study of Low-Energy D-D/D-Be Reaction Accelerator Neutron Source}.
\newblock PhD thesis, Lanzhou University, Lanzhou, China, 2021.
\newblock Ph.D. thesis.

\bibitem{SEO2011S108}
Hee Seo, Jin~Hyung Park, Chan~Hyeong Kim, Ju~Hahn Lee, Chun~Sik Lee, and Jae {Sung Lee}.
\newblock Compton-edge-based energy calibration of double-sided silicon strip detectors in compton camera.
\newblock {\em Nuclear Instruments and Methods in Physics Research Section A: Accelerators, Spectrometers, Detectors and Associated Equipment}, 633:S108--S110, 2011.
\newblock 11th International Workshop on Radiation Imaging Detectors (IWORID).

\bibitem{Lecoq2017}
Paul Lecoq, Alexander Gektin, and Mikhail Korzhik.
\newblock {\em Scintillation and Inorganic Scintillators}, pages 1--41.
\newblock Springer International Publishing, Cham, 2017.

\bibitem{Knoll2010Radiation}
Knoll and F~Glenn.
\newblock Radiation detection and measurement.
\newblock {\em Proceedings of the IEEE}, 69(4):495--495, 2010.

\bibitem{Paatero1974-PAADIC}
P.~Paatero, S.~Manninen, and T.~Paakkari.
\newblock Deconvolution in compton profile measurements.
\newblock {\em Philosophical Magazine}, 30(6):1281--1294, 1974.

\end{thebibliography}

\end{document}